# On the interaction between a current density and a vector potential: Ampère force, Helmholtz tension and Larmor torque.

Germain ROUSSEAUX.

Université de Nice-Sophia Antipolis, INLN – UMR 6638 CNRS-UNICE, 1361 routes des Lucioles, 06560 Valbonne, France.

**Abstract.** Several mathematical formulae are used nowadays in order to compute a magnetic torque. We demonstrate that its more general expression is the vectorial product of the current density with the vector potential. We associate this Larmor's torque with Ampère's force and more specifically with Helmholtz mechanical tension, which is at the origin of the longitudinal stresses in "open" circuits carrying current. We show that Ampère's force enters into the realm of Newtonian Electrodynamics and we explain the absence of contradiction with special relativity. Hence, we provide for the first time a theoretical basis for the numerous experiments, which claimed to have demonstrated the existence of the longitudinal mechanical tension starting with the historical Ampère's hairpin demonstration and the more modern ones of the Graneaus and of Saumont.

Pacs Number : 03.50.De ; 41.20.-q.

*It is true that Ampère's formula is no more admissible today, because it is based on the Newtonian idea of instantaneous action at a distance and it leads notably to the strange consequence that two consecutive elements of the same current should repel each other. Ampère presumed to have demonstrated experimentally this repulsion force, but on this point he was wrong. The modern method, the more rational in order to establish the existence of electrodynamics forces and to determine their value consists in starting from the electrostatic interaction law of Coulomb between two charges (two electrons), whose one of them is at rest in the adopted frame of reference and studying how the interaction forces transform when one goes, thanks to the Lorentz-Einstein relations, to a system of coordinates in which both charges are in motion. One sees the appearance of additional forces proportional to $e^2/c^2$, e being the electrostatic charge and c the light velocity, hence one sees that not only the spin but also the magnetic moment of the electron are of relativistic origin – as Dirac has shown – but that the whole of electromagnetic forces has such an origin.*

<div align="right">*Alfred Kastler, 1977.*</div>

## Introduction

    The forces on conductors are usually described by the Maxwell stress tensor, which leads to a useful interpretation of the forces in terms of the tension along magnetic and electric field lines and the pressure across them.

    The mechanical torque exerted by a magnetic induction on a coil carrying a steady or a quasi-static current can be calculated by taking the vectorial product of the equivalent magnetic moment of the coil with the magnetic induction of say, a magnet [1, 2]. Other expressions appear in the literature as for example the vectorial product of the position vector with the Lorentz force. We will show that all these expressions derived from a general one depending on the vectorial product of the current density with the vector potential.

As the subject is an old one, we have been forced to recall its historical development. However, our findings will give a theoretical basis for the interpretation of modern experiments displaying effects outside the scope of the current understanding/interpretation of electrodynamics. The so-called Larmor torque will allow us to introduce easily the so-called Helmholtz mechanical tension which is closely related to the Ampère force and which explains the appearance of longitudinal stresses in circuits carrying currents whose existence is the subject of a long-standing debate since the original Hairpin experiment designed by Ampère was carried out almost two centuries ago. In addition, we will underline the absence of contradiction between Ampère's law and special relativity suggested by the above quotation of Alfred Kastler thanks to the formalism of Galilean Electromagnetism. Usefulness of this concept for light angular momentum and magnetohydrodynamic dynamo is outlined.

**The Magnetic Torque**

A- The Amperian formulation according to Neumann

How do we compute a magnetic torque? The classical interpretation [1, 2] relies on the use of the Laplace/Lorentz/Grassmann force (denoted only by Lorentz in the following) $\mathbf{F} = \iiint_D \mathbf{j} \times \mathbf{B} d\tau$, which can be written for a closed circuit as $\mathbf{F} = \oint_C I d\mathbf{l} \times \mathbf{B}$. The magnetic induction is calculated with the Biot-Savart formula in the quasi-static limit:

$$\mathbf{B} = \frac{\mu_0}{4\pi} \iiint_{D'} \frac{\mathbf{j} \times \mathbf{r}}{r^3} d\tau'$$

This leads to an interaction force between two closed circuits a and b:

$$\mathbf{F}_{ab} = -\frac{\mu_0}{4\pi} I_a I_b \oint_a \oint_b \mathbf{r}_{ab} \frac{d\mathbf{l}_a . d\mathbf{l}_b}{r_{ab}^3} = -\nabla_a P_{ab}$$

This force is equal to the gradient of the so-called « interaction potential » introduced by F.E. Neumann [3]. The potential is the product of the current inside the b circuit with the flux of the magnetic induction coming from a and going through b:

$$P_{ab} = I_a I_b \frac{\mu_0}{4\pi} \oint_a \oint_b \frac{d\mathbf{l}_a . d\mathbf{l}_b}{r} = \Phi_{ab} I_b$$

We introduce now the mutual inductance, which is a function of the magnetic permeability and of the geometries of both circuits:

$$M_{ab} = \frac{\mu_0}{4\pi} \oint_a \oint_b \frac{d\mathbf{l}_a . d\mathbf{l}_b}{r} = M_{ba}$$

Hence, the interaction force between two circuits can be expressed in the following way:

$$\mathbf{F}_{ab} = -\nabla_a P_{ab} = \nabla_b P_{ab} = \nabla_b (M_{ab} I_a I_b) = \nabla_b (\Phi_{ab} I_b) = -\frac{\mu_0}{4\pi} I_a I_b \oint_a \oint_b \mathbf{r}_{ab} \frac{d\mathbf{l}_a . d\mathbf{l}_b}{r_{ab}^3}$$

with: $r_{ab}^2 = (x_b - x_a)^2 + (y_b - y_a)^2 + (z_b - z_a)^2$ and $\nabla_a r_{ab} = -\nabla_b r_{ab}$.

Moreover, by a dimensional argument: *energy* ≈ *force* × *length* ≈ *torque* × *angle*.

One deduces the existence of an interaction magnetic torque, which is a function of the magnetic flux:

$$C_{ab} = \frac{\partial}{\partial \theta}(P_{ab}) = \frac{\partial}{\partial \theta}(\Phi_{ab} I_b) \text{ with } \Phi_{ab} = B_a S_b \cos\theta$$

such that: $C_{ab} = -I_b B_a S_b \sin\theta$.

A more direct manner is to assess the torque with the Lorentz force:
$$C = \iiint_D \mathbf{r} \times (\mathbf{j} \times \mathbf{B}) d\tau$$
The last formula corresponds to the modern habit.

  B- The Lagrangian formulation according to Maxwell

Maxwell showed that the Neumann's interaction potential can be expressed as a function of the vector potential instead of the magnetic induction [4, 5, 6]:
$$\mathbf{A} = \frac{\mu_0}{4\pi} \iiint_{D'} \frac{\mathbf{j}}{r} d\tau' \text{ or for a closed circuit } \mathbf{A} = \frac{\mu_0}{4\pi} \oint_{C'} \frac{Id\mathbf{l}'}{r} \text{ with } \mathbf{B} = \nabla \times \mathbf{A}$$
Hence, the Neumann's potential is, as a consequence, the product of the current inside the circuit b with the circulation of the vector potential of the circuit a within the circuit b:
$$\Phi_{ab} = \oint_b \mathbf{A}_a . d\mathbf{l}_b = M_{ab} I_a.$$
Starting with the expression of the interaction potential, Maxwell introduced the following expression for the magnetic energy of the whole system, which is analogous to a mechanical kinetic energy:
$$E_m = \frac{1}{2} \iiint_D \mathbf{j}.\mathbf{A} d\tau = \frac{1}{2} L_{aa} I_a^2 + M_{ab} I_a I_b + \frac{1}{2} L_{bb} I_b^2 = \mathrm{L}_m = \sum_{k=1}^n \frac{1}{2} p_k \dot{q}_k$$
with the "generalized velocities" $\dot{q}_k = I_k$, which Maxwell identified to the current intensities and the generalized momenta ("electrokinetic momenta") $p_k = \frac{d\mathrm{L}_m}{d\dot{q}_k} = \sum_{l=1}^n L_{kl} \dot{q}_l$, which Maxwell associated to the vector potentials circulations (the vector potential is interpreted as a linear density of generalized electromagnetic momentum). Following Maxwell [4, 5, 6], $L_{kk}$ is called an "inertia moment" that is an auto-inductance and $L_{kl}$, an "inertia product" that is a mutual inductance. These moments are function of the vacuum permeability, which is a kind of electromagnetic mass or inertia [7]. Maxwell distinguishes the electric variables $q_k$ and the position variables $x_k$ (a length or an angle). The inertia moments $L_{kk} = f(x_k)$ and the inertia products $L_{kl} = g(x_k, x_l)$ are functions only of the latter.

From the electrokinetic Lagrangian $\mathrm{L}_m$, one deduces, by applying the Euler-Lagrange formula (the prime signifies that the force is furnished from the exterior to the system) either:
$$X_i' = \frac{d}{dt} \frac{d\mathrm{L}_{ab}}{d\dot{x}_i} - \frac{d\mathrm{L}_{ab}}{dx_i} = -\frac{d\mathrm{L}_{ab}}{dx_i} = -X_i \text{ i.e. } \mathbf{F}_{ab} = -\nabla \mathrm{L}_{ab} = -\nabla P_{ab}$$
the electromagnetic force (when the generalized coordinate is a position variable) because the interaction Lagrangian does not depend on the velocity; or the electromotive force [3, 4, 5]:
$$Y' = \frac{dp}{dt} - \frac{d\mathrm{L}}{dy} = \frac{dp}{dt} = -Y$$
if the generalized coordinate is an electric variable: $Y = e = -\frac{dp}{dt} = -\frac{d}{dt} \oint_C \mathbf{A}.d\mathbf{l} = -\frac{d\Phi}{dt}$
as the Lagrangian does not depend explicitly on the charges $y = q_k = \int \dot{q}_k dt = \int I_k dt$.

Rare are the authors who have used the Lagrangian formulation of Electrodynamics introduced by Maxwell [8, 9, 10, 11, 12]. According to Henri Poincaré : *"We touch here, as I believe, to the very thinking of Maxwell"* [8]. It seems that Louis De Broglie discovered independently this formulation without reference to Maxwell previous investigations [10, 11].

### C- The Tensorial formulation following Henriot, Costa de Beauregard and Reulos

Émile Henriot then René Reulos introduced what they called a "potential-current" interaction tensor in order to describe the radiation torque experimentally observed by Richard Beth operating with circularly polarized light [13, 14].

1. The "field-polarization" interaction tensor

We recall in this part the very elegant and straightforward demonstration of Reulos [14]. The interaction between an electric dipole and an electric field leads to the existence of not only an energy but also a torque:

$$\frac{dE_p}{d\tau} = -\mathbf{P}.\mathbf{E} \text{ and } \frac{d\mathbf{C}_p}{d\tau} = \mathbf{P} \times \mathbf{E}$$

Reulos noticed that one can condense these two equations by introducing the so-called "field-polarization" interaction tensor whose expression is the product of an antisymmetric tensor built with the electric field by another antisymmetric tensor built with the electric polarization [14] :

$$d\begin{vmatrix} E_p^1 & C_p^3 & -C_p^2 & C_p^1 \\ -C_p^3 & E_p^2 & C_p^1 & C_p^2 \\ C_p^2 & -C_p^1 & E_p^3 & C_p^3 \\ -C_p^1 & -C_p^2 & -C_p^3 & E_p^4 \end{vmatrix} = \begin{vmatrix} 0 & \mathbf{E}_p^3 & -\mathbf{E}_p^2 & \mathbf{E}_p^1 \\ -\mathbf{E}_p^3 & 0 & \mathbf{E}_p^1 & \mathbf{E}_p^2 \\ \mathbf{E}_p^2 & -\mathbf{E}_p^1 & 0 & \mathbf{E}_p^3 \\ -\mathbf{E}_p^1 & -\mathbf{E}_p^2 & -\mathbf{E}_p^3 & 0 \end{vmatrix} \bullet \begin{vmatrix} 0 & \mathbf{P}_p^3 & -\mathbf{P}_p^2 & \mathbf{P}_p^1 \\ -\mathbf{P}_p^3 & 0 & \mathbf{P}_p^1 & \mathbf{P}_p^2 \\ \mathbf{P}_p^2 & -\mathbf{P}_p^1 & 0 & \mathbf{P}_p^3 \\ -\mathbf{P}_p^1 & -\mathbf{P}_p^2 & -\mathbf{P}_p^3 & 0 \end{vmatrix} d\tau$$

The interaction between a magnetic moment and a magnetic field is similar:

$$\frac{dE_a}{d\tau} = -\mathbf{M}.\mathbf{B} \text{ and } \frac{d\mathbf{C}_a}{d\tau} = \mathbf{M} \times \mathbf{B}$$

and one can use now the tensorial notation of Reulos:

$$d\begin{vmatrix} E_a^1 & C_a^3 & -C_a^2 & C_a^1 \\ -C_a^3 & E_a^2 & C_a^1 & C_a^2 \\ C_a^2 & -C_a^1 & E_a^3 & C_a^3 \\ -C_a^1 & -C_a^2 & -C_a^3 & E_a^4 \end{vmatrix} = \begin{vmatrix} 0 & \mathbf{B}_a^3 & -\mathbf{B}_a^2 & \mathbf{B}_a^1 \\ -\mathbf{B}_a^3 & 0 & \mathbf{B}_a^1 & \mathbf{B}_a^2 \\ \mathbf{B}_a^2 & -\mathbf{B}_a^1 & 0 & \mathbf{B}_a^3 \\ -\mathbf{B}_a^1 & -\mathbf{B}_a^2 & -\mathbf{B}_a^3 & 0 \end{vmatrix} \bullet \begin{vmatrix} 0 & \mathbf{M}_a^3 & -\mathbf{M}_a^2 & \mathbf{M}_a^1 \\ -\mathbf{M}_a^3 & 0 & \mathbf{M}_a^1 & \mathbf{M}_a^2 \\ \mathbf{M}_a^2 & -\mathbf{M}_a^1 & 0 & \mathbf{M}_a^3 \\ -\mathbf{M}_a^1 & -\mathbf{M}_a^2 & -\mathbf{M}_a^3 & 0 \end{vmatrix} d\tau$$

2. The "potential-current" interaction tensor

René Reulos based his reasoning on the fact that the four-potential and the four-current interact in a similar way with respect to energy [14]:

$$\frac{dE_{charge}}{d\tau} = \rho V = -j_4 A_4 \text{ and } \frac{dE_{courant}}{d\tau} = -\mathbf{j}.\mathbf{A} = -\sum_{k=1}^{3} j_k A_k$$

that is for the total energy:

$$dE = (\rho V - j_1 A^1 - j_2 A^2 - j_3 A^3) d\tau = -j_\mu A^\mu d\tau = dC_4$$

Hence, Reulos built potentials and currents tensor whose product (as a matrix product) is by definition the "potential-current" interaction tensor:

$$d\begin{vmatrix} E & C_3 & -C_2 & C_1 \\ -C_3 & E & C_1 & C_2 \\ C_2 & -C_1 & E & C_3 \\ -C_1 & -C_2 & -C_3 & E \end{vmatrix} = \begin{vmatrix} iV & A_3 & -A_2 & A_1 \\ -A_3 & iV & A_1 & A_2 \\ A_2 & -A_1 & iV & A_3 \\ -A_1 & -A_2 & -A_3 & iV \end{vmatrix} \bullet \begin{vmatrix} i\rho & j_3 & -j_2 & j_1 \\ -j_3 & i\rho & j_1 & j_2 \\ j_2 & -j_1 & i\rho & j_3 \\ -j_1 & -j_2 & -j_3 & i\rho \end{vmatrix} d\tau$$

and whose spatial components define a density of magnetic torque per unit volume:

$$\begin{cases} dC_1 = (j_2 A_3 - j_3 A_2) d\tau \\ dC_2 = (j_3 A_1 - j_1 A_3) d\tau \\ dC_3 = (j_1 A_2 - j_2 A_1) d\tau \end{cases} \text{ that is } d\mathbf{C} = \mathbf{j} \times \mathbf{A} d\tau$$

However, Beth's experiments with the dielectrics are explained thanks to the torque $d\mathbf{C}_p = \mathbf{P} \times \mathbf{E} d\tau$. Nevertheless, Reulos noticed the role played by the vector $d\mathbf{C} = \mathbf{j} \times \mathbf{A} d\tau$ in order to demonstrate the origin of the torque, which appears on a "Hertzian polarizer/analyzer" constituted with vertical metallic rods suspended to a frame and submitted to an electromagnetic wave [14]. Besides, by expressing the torque density on a length element in function of the current and by introducing the integral expression of the vector potential in function of the current, Reulos has shown that the integral torque vector exerted by a circuit a on another circuit b can rearrange in the following form [15, 16] :

$$\mathbf{C}_{ab} = -I_b \oint_b \mathbf{A}_a \times d\mathbf{l}_b = -I_a I_b \oint_a \oint_b \frac{d\mathbf{l}_a \times d\mathbf{l}_b}{r}$$

Starting from the infinitesimal expression of the magnetic energy $W = I\Phi$, Olivier Costa de Beauregard noticed that it could be reorganized ($dW = I\mathbf{A}.d\mathbf{l} = \mathbf{T}.d\mathbf{l}$) in order to display the Ampère's tension (as he called it) $\mathbf{T} = I\mathbf{A}$, which is at the origin of an infinitesimal torque $d\mathbf{C} = d\mathbf{l} \times \mathbf{T} = Id\mathbf{l} \times \mathbf{A}$ and, which is nothing but the expression given by Reulos [17].

3. The electromagnetic stress tensor of Reulos and Costa de Beauregard

Besides, O. Costa de Beauregard introduced the "elastic" tensor $N^{kl} = A^k J^l - \frac{1}{2} A^i J_i \delta^{kl}$, which constitutes an alternative to the use of the Minkowski's tensor (function of the fields) in order to express the Lorentz force $K_i = F_{ik} J^k = -\partial_k (F_{il} F^{il} - 1/2 F_{rs} F^{rs} \delta_i^k)$ (where $F_{ik} = \partial_i A_k - \partial_k A_i$ is the Faraday/Maxwell/Minkowski tensor) in a four-dimensional manner and only in terms of the four-potentials [18]. The spatial part of the Costa De Beauregard tensor $v^{kl} = N^{lk} - N^{kl}$ is equal to the opposite of the density per unit volume of the magnetic torque $d\mathbf{C} = \mathbf{j} \times \mathbf{A} d\tau$. Finally, R. Reulos proposed, following Costa de Beauregard, a variational method in order to explain the origin of the energy tensor [16, 19].

D-The thermodynamic formulation according to Laue and De Haas

Max Von Laue was the doctoral student of Max Planck who was a specialist of thermodynamics. Indeed, Planck was the first to treat the extension of relativity to thermodynamics. Hence, it is not surprising that his student used thermodynamics in order to tackle the problem of formulating an electromagnetic stress-energy tensor with the goal to deduce the expression for the magnetic torque [20, 21].

As a matter of fact, the first principle of thermodynamics can be expressed in the following form for a constant number of particles and volume: $d\varepsilon_0 = T_0 ds$. A Galilean change of frame of reference modifies the rest energy density by adding a kinetic energy density $\varepsilon = \varepsilon_0 + \frac{1}{2}\rho v^2 = \varepsilon_0 + \frac{g^2}{2\rho}$ where we defined the impulsion density: $\mathbf{g} = \rho \mathbf{v}$. From it, we deduce the expression of the first principle in a moving frame of reference: $d\varepsilon = Tds + \mathbf{v}.d\mathbf{g}$ with $T = T_0$. Hence, we can define a generalized thermodynamic potential, which is a minimum at equilibrium: such that $df = Tds - \mathbf{g}.d\mathbf{v}$. The product $\mathbf{v}.\mathbf{g}$ appears as the kinetic contribution to the energy density. Besides, we know that the generalized "impulsion" of a massive charge particle in presence of a vector potential writes: $\mathbf{p} = m\mathbf{v} + q\mathbf{A}$. Following Maxwell, one can define an "electro-tonic impulsion" density $\mathbf{G} = \rho_e \mathbf{A}$, which is the product of the electric charge density with the vector potential, which physically is an electromagnetic momentum per unit charge. Now, the product of the charges velocity by the "electro-tonic" momentum is an energy: $T_{em} = \mathbf{v}_e.\mathbf{G} = \rho_e \mathbf{v}_e.\mathbf{A} = \mathbf{j}.\mathbf{A}$ [20, 21]. The quadri-dimensional generalization is straightforward: $T^{\mu\nu}_{Laue} = V^{\mu}G^{\nu} = j^{\mu}A^{\nu} = T^{\mu\nu}_{ja}$. We are willing to construct an "impulsion-torque tensor" from the stress-energy tensor $T^{\mu\nu}_{ja} = j^{\mu}A^{\nu}$ attributed to Gustav Mie (1912-1913) by De Haas [21]. We recall that in hydrodynamics, one can construct a rotation tensor from the velocity tensor with indicial components $\frac{\partial u_i}{\partial x_j}$ thanks to anti-symmetrisation: $\mathbf{\Omega}_{ij} = \left(\frac{\partial u_i}{\partial x_j} - \frac{\partial u_j}{\partial x_i}\right)$. We suppose that the impulsion-torque tensor has the following antisymmetric form: $N^{\mu\nu}_{ja} = j^{\mu}A^{\nu} - j^{\nu}A^{\mu}$ whose spatial components correspond to the volume density of the magnetic torque: $\mathbf{n} = \mathbf{j} \times \mathbf{A}$ [21]. Paul De Haas has recently re-examined the problem of energy conservation in relativity by following the path of Von Laue by exploring the conservation equation $\partial_{\mu}(j^{\mu}A^{\nu}) = 0$, which according to him, leads to the removal of several paradoxical problems in Electrodynamics [21].

E-Larmor's precession

The Lorentz force can be rewritten as a function of the potentials:
$$\frac{d}{dt}(m\mathbf{v} + q\mathbf{A}) = -q\nabla(V - \mathbf{v}.\mathbf{A})$$
with:
$$\frac{d\mathbf{A}}{dt} = \frac{\partial \mathbf{A}}{\partial t} + (\mathbf{v}.\nabla)\mathbf{A}$$
where $\nabla$ applies *only* to $\mathbf{A}(\mathbf{r},t)$ and not to $\mathbf{v} = \frac{d\mathbf{r}}{dt}$.

The generalized impulsion is by definition $\mathbf{P} = m\mathbf{v} + q\mathbf{A}$. The angular momentum associated to it writes $\mathbf{L} = \mathbf{r} \times \mathbf{P} = \mathbf{r} \times (m\mathbf{v} + q\mathbf{A}) = \mathbf{l} + \mathbf{r} \times q\mathbf{A}$. One deduces the balance for the angular momentum:
$$\frac{d\mathbf{L}}{dt} = q\mathbf{v} \times \mathbf{A} + q\mathbf{r} \times \nabla(\mathbf{v}.\mathbf{A} - V)$$

If $V = V(r)$ as for an electron in an atom submitted to a uniform magnetic field whose vector potential is $\mathbf{A} = \frac{1}{2}\mathbf{B} \times \mathbf{r}$ ($\nabla.\mathbf{A} = 0$), then we get:

$$\frac{d\mathbf{L}}{dt} = q\mathbf{v} \times \mathbf{A} = \frac{q}{2m}\mathbf{l} \times \mathbf{B}$$

Usually, the diamagnetic contribution $q\mathbf{A}$ is negligible in $\mathbf{L}$, so we end up with:

$$\frac{d\mathbf{l}}{dt} \approx q\mathbf{v} \times \mathbf{A} = \frac{q}{2m}\mathbf{l} \times \mathbf{B} = \frac{e}{2m}\mathbf{B} \times \mathbf{l} = \mathbf{\Omega} \times \mathbf{l}$$

where $\mathbf{\Omega} = \frac{e}{2m}\mathbf{B}$ is the well-known Larmor angular velocity.

As a partial conclusion, the famous Lamor precession is a direct consequence of the existence of the magnetic torque :

$$q\mathbf{v} \times \mathbf{A} \approx \rho_e d\tau \mathbf{v} \times \mathbf{A} = \rho_e \mathbf{v} \times \mathbf{A} d\tau = \mathbf{J} \times \mathbf{A} d\tau.$$

**Links between the different formulations**

If we extend Maxwell's reasoning, a torque is a generalized force that one can obtain by using the Euler-Lagrange equation when applied to an angle from the interaction Lagrangian [6] :

$$C_{ab} = \frac{\partial}{\partial \theta}(\mathrm{L}_{ab}) = \frac{\partial}{\partial \theta}(\Phi_{ab}I_b)$$

with $\Phi_{ab} = \oint_b \mathbf{A}_a . d\mathbf{l}_b = \mathbf{B}_a \cdot \mathbf{S}_b = B_a S_b \cos\theta$. We find the torque derived previously from the Amperian procedure (here, $\mathbf{n} = \mathbf{S}_b / S_b$ denotes the unit vector perpendicular to the coil b):

$$\mathbf{C}_{ab} = -I_b B_a S_b \sin\theta \mathbf{n} = -I_b \mathbf{n} \oint_b \mathbf{A}_a . d\mathbf{l}_b \frac{\sin\theta}{\cos\theta} = \oint_b I_b d\mathbf{l}_b \times \mathbf{A}_a = \iiint_b \mathbf{j}_b \times \mathbf{A}_a d\tau$$

which displays, after some modifications, the following magnetic torque density:

$$\frac{d\mathbf{C}_{ab}}{d\tau} = \mathbf{j}_b \times \mathbf{A}_a$$

As an example of the equivalence between all formulations, we will calculate the magnetic torque for the well-known Pellat electrodynamometer. Given a fixed horizontal solenoid of whatever section and characterised by $n_s$ coils per unit length. We put a coil constituted with several enrolments of section $S_b$ inside the solenoid. This one is carrying a current $I_s$ whereas the intensity in the coil is $I_b$. The calculus of the force and the torque exerted by the solenoid on the inner coil implies the evaluation of the mutual inductance $M_{sb}$. Besides, the mechanical action reduces to a torque as the variation of the mutual inductance for a horizontal translation is null as one assumes the solenoid to be infinite. Due to symmetry, the torque's axis is perpendicular to the plane enclosing the coil surfaces. One has :

$$M_{sb} = \frac{\Phi_{sb}}{I_s} = \frac{1}{I_s}\oint_b \mathbf{A}_s . d\mathbf{l} = \frac{B_s S_b \cos\theta}{I_s} = \mu_0 n_s S_b \cos\theta$$

The inner coil is submitted to the following torque, following Neumann:

$$C_{sb} = I_s I_b \frac{\partial M_{sb}}{\partial \theta} = -I_s I_b \mu_0 n_s S_b \sin\theta = m_b B_s \sin(-\theta)$$

However, the inner coil is equivalent to a magnet whose magnetic moment is $m_b = SI_b$ and is submitted to the usual magnetic induction created by a solenoid $B_s = \mu_0 n_s I_s$. Hence, the inner coil is submitted to the well-known torque: $\mathbf{C}_{sb} = \mathbf{m}_b \times \mathbf{B}_s$. If the coil is placed outside the solenoid, no torque is detected as the product of the current by the magnetic flux is constant.

# The Longitudinal Tension

H. von Helmholtz introduced in 1870 the concept of "longitudinal tension" between two successive current elements [4 Chapter 4, 4.1 to 4.3] : an "open" linear current element is submitted to two forces $-I\mathbf{A}(\mathbf{r}_1)$ and $I\mathbf{A}(\mathbf{r}_2)$ between its two ends $\mathbf{r}_1$ and $\mathbf{r}_2$ due to the presence of a space-dependent vector potential. Maxwell in 1873 derives independently in his Treatise the mathematical expression for the Helmholtz mechanical tension but without discussing its physical implication as Helmholtz [6]. We sketched here the simpler derivation of Maxwell. One recalls first that the product of the current with the circulation of the vector potential stands for the magnetic energy. Maxwell used variational calculus in order to derive the force on a current element by differentiating with respect to a virtual displacement $\delta\mathbf{l}$ the magnetic energy:

$$\int d\mathbf{f}.\delta\mathbf{l} = \delta\left(I\int \mathbf{A}.d\mathbf{l}\right) = I\left(\int \delta\mathbf{A}.d\mathbf{l}\right) + I\left(\int \mathbf{A}.d\delta\mathbf{l}\right) = I\int (\delta\mathbf{l}.\nabla)\mathbf{A}.d\mathbf{l} + I\int \mathbf{A}.\delta\mathbf{l} - I\int \delta\mathbf{l}.(d\mathbf{l}.\nabla)\mathbf{A}$$

The first and third term lead to the Lorentz force whereas the second term is the Helmholtz tension and integrates to zero if the circuit is closed.

Joseph Larmor provides in 1897 several mathematical demonstrations for the existence of a longitudinal tension but with no references to Maxwell and Helmholtz's previous demonstrations. We reproduce Larmor's main reasoning for two demonstrations with his own words and some comments [4 Appendix 9, 22 III, 23 p. 223-226, 24, 25]:

*« [For linear conductors,] the total electrokinetic energy is :*

$$\iint Mi ds i' ds' \text{ where } M = \frac{1}{r}\cos(ds, ds') + \frac{1}{2}\frac{\partial^2 r}{\partial s \partial s'}$$

*If the currents are uniform all along the linear conductors, the second term in M integrates to nothing when the circuit are complete, and we are thus left with the Ampère-Neumann expression for the total energy of the complete current, from which the Amperan law of force may be derived in the known manner by the method of variations. But it must be observed that, as the localization of the energy is in that process neglected, the legitimate result is that the forcive of Ampère, together with internal stress as yet undetermined between contiguous parts of the conductors, constitute the total electromagnetic forcive : it would not be justifiable to calculate the circumstances of internal mechanical equilibrium from the Amperean forcive alone, unless the circuits are rigid. For example, if we suppose that the circuits are perfectly flexible, we may calculate the tension in each, in the manner of Lagrange, by introducing into the equation of variation the condition of inextensibility. We arrive at a tension $i \int Mi' ds'$ where i is the current at the place considered; whereas the tension as calculated from Ampère's formula for the forcive would in fact be constant, the forcive on each element of the conductor being wholly at right angles to it. »*

*« Consider [now] a current system to be built up of physical current elements of the form $(j_x, j_y, j_z)d\tau$, the energy associated with an element of volume $d\tau$, as existing in the surrounding field and controlled by the element, is :*

$$T_L = \mathbf{j}.\mathbf{A} d\tau$$

*The ponderomotive force acting on the element will be derived from a potential energy function – $T_L$, by varying the coordinate of the material framework: it must in fact consist, per unit volume of a force :*

$$\left( j_x \frac{\partial A_x}{\partial x} + j_y \frac{\partial A_y}{\partial x} + j_z \frac{\partial A_z}{\partial x}, j_x \frac{\partial A_x}{\partial y} + j_y \frac{\partial A_y}{\partial y} + j_z \frac{\partial A_z}{\partial y}, j_x \frac{\partial A_x}{\partial z} + j_y \frac{\partial A_y}{\partial z} + j_z \frac{\partial A_z}{\partial z} \right)$$

*and a couple :*

$$\mathbf{j} \times \mathbf{A}$$

*The former being derived from a translational, the latter from a rotational virtual displacement of the element. »*

« *The traction in the direction of the current would introduce an additional tension, equal to the current multiplied by the component of the vector potential in its direction, which is not usually constant along the circuit, and so may be made the subject of experimental test with liquid conductors, as it would introduce differences of fluid pressure. There will also be an additional transverse shearing stress, which should reveal itself in experiments on solid conductors with sliding contacts.* »

Darrigol has pointed out correctly that the derivation of the Larmor's force was flawed [4 Appendix 9]. Indeed, Larmor treated wrongly the current density as a "force" and not a "density" according to Maxwell terminology [6]. Darrigol did not comment on the torque calculation. Hence, Larmor derived one tension, one force and a torque. We believe the tension to be the one of Helmholtz and the torque will be called by us the Larmor torque because its expression as a cross-product was first derived by him before Reulos, Costa De Beauregard, Carpenter and De Haas. Finally, Larmor points out that the use of mercury will certainly reveals the longitudinal stresses…

**The Force of Ampère**

As recalled beautifully by Kastler, the main scientific achievement of Ampère was to derive the force between two current elements $I_a d\mathbf{l}_a$ and $I_b d\mathbf{l}_b$ separated by a distance $r$ [26, 27, 28]:

$$d^2\mathbf{F}_A = -\frac{\mu_0}{4\pi} I_a I_b \frac{1}{r^2} \left[ 2\frac{\mathbf{r}}{r}(d\mathbf{l}_a . d\mathbf{l}_b) - 3\frac{\mathbf{r}}{r^3}(d\mathbf{l}_a . \mathbf{r})(d\mathbf{l}_b . \mathbf{r}) \right]$$

Its differs from Laplace/Lorentz/Grassmann's law used in modern textbooks according to [29]:

$$d^2\mathbf{F}_A = I_a d\mathbf{l}_a \times \frac{\mu_0}{4\pi} \frac{I_b d\mathbf{l}_b \times \mathbf{r}}{r^3} + d\mathbf{l}_b . \nabla_b \left( \frac{\mu_0}{4\pi} \frac{I_a I_b \mathbf{r}.(\mathbf{r}.d\mathbf{l}_b)}{r^3} \right)$$

For closed circuits, both forces give the same predictions. However, Moon and Spencer have examined very deeply in a series of articles [30, 31, 32] the various possible force laws in accordance with Ampère's experiments. As a conclusion, they rejected forcefully Grassmann's law, which for instance is not in agreement with Newton's third law…

The forces on infinitesimal circuit elements are usually given by the Lorentz force law, given the local field due to the entire circuit. However, while the fields given by the Biot-Savart law are written as the sum over individual infinitesimal elements, only the sum is assumed to be correct. Moreover, an individual infinitesimal circuit element is interpreted as a moving charge, or a current that is not divergence-free and therefore not steady, so the fields are not correctly given by magnetostatics or magnetoquasistatic.

We insist on the fact that the law of Ampère is experimental so it cannot be questionable (contrary to A. Kastler's introducing quotation) as soon as the conditions of the experimental realizations are known. From its basic expression, it appears clearly that the law could be (wrongly) interpreted as an action at distance since it features the two current elements and the separation between them. However, the Ampère's force can be rewritten in such a way that one current element interacts with the vector potential (or alternatively the magnetic field) created by the other current element [33]. So action at distance is replaced by the interaction between a source and a field created by a distant source, which propagates instantaneously. Now, the Ampère's force satisfies to Newton's third law. Moreover, the equality of action and reaction is only compatible with instantaneous interactions otherwise the simultaneity will be relative as demonstrated by H. Poincaré [34, 35]. This last fact is one of the major arguments used by the opponents of the longitudinal stresses as implied by Ampère's law because it is in contradiction with special relativity. Indeed, special relativity and hence the full set of Maxwell equations cannot be made compatible with Ampère's law. However, what has not been noticed before is the fact that Ampère's law is compatible with the so-called Galilean magnetic limit of Maxwell equations as one deals with divergenceless currents in the quasi-stationary approximation [36, 37]. We underline strongly that a current, which is divergence-free (magnetic limit), is not necessarily steady as often assumed. Hence, Ampère's law (so longitudinal stresses) as an approximation is completely compatible with Galilean electromagnetism and so with the principle of relativity [36, 37]. Of course, for rapid oscillations of the currents, Ampère's law will break down (waves will propagate) and is indeed incompatible with special relativity and more precisely with the light velocity postulate but still not with the relativity postulate. That's why we think that the Graneaus (father and son) were particularly right to call their book Newtonian Electrodynamics [38] as Ampère's law is strictly valid only in the realm of Galilean Physics...

Following Cornille, Ampère's force for volume elements $d^3\tau$ becomes [33]:

$$d^6\mathbf{F}_A = -\frac{\mu_0}{4\pi}\frac{1}{r^2}\left[2(\mathbf{J}_a.\mathbf{J}_b) - 3\left(\mathbf{J}_a.\frac{\mathbf{r}}{r}\right)\left(\mathbf{J}_b.\frac{\mathbf{r}}{r}\right)\right]\frac{\mathbf{r}}{r}d^3\tau_a d^3\tau_b$$

When $\mathbf{J}_a$ and $\mathbf{J}_b$ are co-linear, Ampère predicted the existence of a repulsion:

$$d^6F_A = \frac{\mu_0}{4\pi}\frac{J_a J_b}{r^2}d^3\tau_a d^3\tau_b$$

which is in contradiction with Laplace/Lorentz/Grassmann law, which predicts a zero force between both current elements. Hence, one recovers roughly the mechanical tension introduced by Helmholtz from Ampère's force in the longitudinal direction:

$$T_H \propto \frac{\mu_0}{4\pi}\frac{J_a}{r}d^3\tau_a.\frac{J_b d^3\tau_b}{r} \propto dA_a.I_b$$

where $dA_a$ is the vector potential created by the current density $J_a$ interacting with the current intensity $I_b$.

With the help of the Swiss physicist De La Rive, Ampère performed the well-known hairpin experiment to confirm his prediction. It consists in connecting two parallel troughs filled with mercury to a battery and completing the circuit by a bridge perpendicular to the two wires forming the hairpin and floating on the mercury. The hairpin is isolated except at its extremities where the repulsion forces act and propel the system away from the battery whatever is the sign of the current [26, 27]. Several reproductions have confirmed the existence of the effect whereas the theoretical explanation given by Ampère is still the subject of an intense controversy in the recent literature [38, 39, 40, 41].

H. Helmholtz, his Russian student N. Schiller, the Maxwellians G. F. Fitzgerald and O. J. Lodge have made several attempts to discover experimentally the additional longitudinal stresses derived independently by Maxwell, Helmholtz and Larmor. All the experiments were negative [4 Chapter 6, 22 III, 23 p. 223-226, 24 Two and Three]. Concerning Helmholtz's ones, he expected charges effect to occur sine he wrote the new forcives in function of the time derivative of the charge density thanks to charge conservation: he was wrong (Larmor made the same physical/mathematical error [42]) as the current density is divergenceless within the Galilean magnetic limit [36, 37]. However, none of these authors made the link between the tension $\mathbf{T} = I\mathbf{A}$ and the Hairpin experiment of Ampère…

It seems that Costa de Beauregard was the first to envisage the mathematical relationship more than one century and half after Ampère's experiments [17, 43]. According to us, one of the most striking experimental results concerning the Helmholtz tension is the experiment of Rémi Saumont [44, 45]. Indeed, as we have seen Larmor suggested that liquid metal conductors would certainly reveals differences of fluid pressure due to the predicted mechanical tension. Saumont, without knowing Larmor suggestion, designed a very clever weighing method in order to measure the apparent lightening or increase in weight of a horizontal metallic circuit connected to a balance pan, carrying an electric current and whose vertical ends were plunged in two separate beakers of mercury related to a battery. For example, lightening is obtained when the ends are directed downwards. He has shown by a careful study that the difference in weight is proportional to the square of the current intensity as predicted by the Helmholtz tension/Ampère force [44, 45].

Other experiments have been carried out in order to demonstrate the existence of the Ampère's force. For example, the behaviour of electromagnetic railguns is thought to be explained by the Helmholtz tension/Ampère force as well as the wire exploding phenomena [38, 45, 46]. We hope that our study will close the debate with respect to the existence of the longitudinal stresses [47, 48, 49, 50, 51, 52].

## Conclusions

As a conclusion, we have shown that the expression of Larmor is the more general expression for a magnetic torque whose existence is confirmed by everyday experiments. It is then straightforward to deduce the existence of a Helmholtz mechanical tension, which is closely related to Ampère's force. Much more studies are needed now in order to compute directly the Helmholtz tension for particular case as for Saumont experiments. In addition, we have recalled Reulos reasoning for introducing the Larmor torque. Present days studies of the light angular momentum should make the link with the thought experiment of Reulos with metallic rods where one should observe a rotation of the frame carrying the rods when submitted to an electromagnetic wave [52]. As foreseen by Larmor, the use of liquid metal like in Magnetohydrodynamics is a perfect field of testing for the Helmholtz tension. By the way, it is strange that the long-standing problem of earth dynamo was systematically approached through the use of Lorentz force at the notable exception of the mechanical model of Rikitake who precisely used a formulation in terms of the magnetic torque [53]. Hence, maybe Larmor's torque could give some clues for the generation of motion by a liquid metal carrying induced currents…

# Bibliography

[1] G. Bruhat, Electricité, 8$^{ème}$ edition révisée par G. Goudet, Masson, 1963.

[2] P. Lorrain, D.R. Corson & F. Lorrain, Les Phénomènes Electromagnétiques, Dunod, 2002.

[3] F.E. Neumann, Recherches sur la théorie mathématique de l'induction, Traduction de M.A. Bravais, Journal de Mathématiques Pures et Appliquées, Tome XIII, p. 113-178, Avril 1848. http://gallica.bnf.fr/

[4] O. Darrigol, Electrodynamics from Ampère to Einstein, Oxford University Press, 2000.

[5] J. Clerk Maxwell, A Dynamical Theory of the Electromagnetic Field (1865), W.D. Niven, ed., The Scientific Papers of James Clerk Maxwell, 2 vols., New York, 1890.

[6] J. Clerk Maxwell, A treatise on electricity and magnetism (1873), Volume II, Chapitres III à VIII, Dover Publications, 1954. http://gallica.bnf.fr/

[7] R. Anderson, On an Early Application of the Concept of Momentum to Electromagnetic Phenomena: The Whewell-Faraday Interchange, Studies in the History and Philosophy of Science, 25, p. 577-594, 1994.

[8] H. Poincaré, Electricité et Optique, La lumière et les théories électrodynamiques, Ed. G. Carré et C. Naud, 1901. http://gallica.bnf.fr/

[9] W.F.G. Swann, Relativity and Electrodynamics, Reviews of Modern Physics, Vol. 2, Num. 3, p. 243-346, 1930.

[10] L. De Broglie, Diverses questions de mécanique et de thermodynamique classiques et relativistes, Cours à l'Institut Henri Poincaré (1948), Chapitre IX, Springer, 1995.

[11] L. De Broglie, Energie Libre et Fonction de Lagrange. Application à l'Electrodynamique et à l'Interaction entre Courants et Aimants Permanents, Portugaliae Physica, Vol. 3, Fasc. 1, p. 1-20, 1949.

[12] D.A. Wells, Lagrangian Dynamics, Chapter 15, Schaum's Outlines, Mc Graw Hill, 1967.

[13] E. Henriot, Les Couples de Radiation et les Moments Electromagnétiques, Mémorial des Sciences Physiques, Fascicule XXX, Gauthier-Villars, 1936.

[14] R. Reulos, Recherches sur la théorie des corpuscules, Annales de l'Institut Fourier, Tome 5, p. 455-568, 1954. http://archive.numdam.org/article/AIF_1954__5__455_0.pdf

[15] R. Reulos, Interactions électromagnétiques et couples de radiation, Helvetica Physica Acta, vol. 27, p. 491-493, 1954.

[16] R. Reulos, L'effet potentiel vecteur (Interaction Courant Potentiel et les Couples de Radiation), Compte rendu des séances de la S.P.H.N. de Genève, NS, vol. 2, fasc. 1, p. 87-96, 1967.

[17] O. Costa de Beauregard, Statics of filaments and magnetostatics of currents : Ampère tension and the vector potential, Physics Letters A, 183, p. 41-42, 1993.

[18] O. Costa de Beauregard, Définition et interprétation d'un nouveau tenseur élastique et d'une nouvelle densité de couple en électromagnétisme des milieux polarisés, Comptes Rendus de l'Académie des Sciences de Paris, 217, p. 662-664, 1943.

[19] R. Reulos, Sur un nouveau tenseur d'énergie, Compte rendu des séances de la S.P.H.N. de Genève, Séance du 4 mars, p. 47-60, 1971.

[20] M. Von Laue, Zur Dynamik der Relativitätstheorie, Annalen der Physik, Vol. 35, p. 524-542, 1911. http://gallica.bnf.fr/

[21] E.P.J. de Haas, A renewed theory of electrodynamics in the framework of Dirac-ether, PIRT Conference, London, September 2004. http://home.tiscali.nl/physis/deHaasPapers/PIRTpaper/deHaasPIRT.html

[22] J. Larmor, A dynamical theory of the electric and luminiferous ether, Part I, II, and III, in Mathematical and Physical Papers, 2 Vol., Cambridge, 1929.

[23] B. Hunt, The Maxwellians, Cornell Paperbacks, Cornell University Press, 2005.

[24] J. Buchwald, The Creation of Scientific Effects : Heinrich Hertz and Electric Waves, The University of Chicago Press, 1994.

[25] O. Darrigol, The electrodynamics of moving bodies from Faraday to Hertz, Centaurus, Vol. 36, p. 245-260, 1993.

[26] A.-M. Ampère, Théorie Mathématique des Phénomènes Electrodynamiques, Blanchard, Paris, 1958.

[27] J.R. Hofmann, Ampère, Electrodynamics, and Experimental Evidence, Osiris, 2$^{nd}$ Series, Vol. 3, p. 45-76, 1987.

[28] A. Kastler, Ampère et les lois de l'électrodynamique, Rev. Hist. Sci., XXX/2, p. 1193-1207, 1977.

[29] O. Darrigol, The electrodynamics revolution in Germany as documented by early german expositions of « Maxwell's theory », Archive for History of Exact Sciences, Vol. 45, Num. 3, p. 189-280, 1993.

[30] P. Moon & D.E. Spencer, The Coulomb force and the Ampère force, J. Franklin Inst., 257, 305, 1954.

[31] P. Moon & D.E. Spencer, Interpretation of the Ampère experiments, J. Franklin Inst., 257, 203, 1954.

[32] P. Moon & D.E. Spencer, A new electrodynamics, J. Franklin Inst., 257 (5), 369, 1954.

[33] P. Cornille, Advanced Electromagnetism and Vacuum Physics, World Scientific, 2003.

[34] O. Darrigol, Henri Poincaré's Criticism of *Fin de Siècle* Electrodynamics, Stud. Hist. Phil. Mod. Phys., Vol. 26, No. 1, p. 1-44, 1995.

[35] G. Granek, Poincaré's Contributions to Relativistic Dynamics, Stud. Hist. Phil. Mod. Phys., Vol. 31, No. 1, p. 15-48, 2000.

[36] M. Le Bellac & J.-M. Lévy-Leblond, Galilean electromagnetism, Nuovo Cimento, 14B, 217-233 (1973).

[37] G. Rousseaux, Lorenz or Coulomb in Galilean Electromagnetism ?, EuroPhysics Letters, 71 (1), p. 15-20, 2005; M. de Montigny & G. Rousseaux, On the electrodynamics of moving bodies at low velocities, European Journal of Physics, 27, p. 755-768, 2006.

[38] P. Graneau & N. Graneau, Newtonian Electrodynamics, World Scientific, 1996.

[39] N. Graneau, T. Phipps Jr, D. Roscoe, An experimental confirmation of longitudinal electrodynamic forces, Eur. Phys. J. D 15, 87, 2001.

[40] A.E. Robson, Comment on : An experimental confirmation of longitudinal electrodynamic forces by N. Graneau, T. Phipps Jr, and D. Roscoe, Eur. Phys. J. D 22, 117–118, 2003.

[41] J.P. Wesley, Proposed motors driven solely by Ampère repulsion, EuroPhysics Letters, 63 (2), p.214-218, 2003.

[42] J. Larmor, On the theory of Electrodynamics, in Mathematical and Physical Papers, 2 Vol., Cambridge, 1929.

[43] O. Costa de Beauregard, Electromagnetic Gauge as Integration Condition : Einstein's Mass-Energy Equivalence Law and Action-Reaction Opposition, in Advanced Electromagnetism, Foudations, Theory and Applications, Ed. T.W. Barrett & D.M. Grimes, p. 77-104, 1995.

[44] R. Saumont, Mechanical effects of an electrical current in conductive media. Experimental investigation of the longitudinal Ampère force, Phys. Lett. A, 165, p. 307, 1992.

[45] R. Saumont, Ampère Force : Experimental Tests, in Advanced Electromagnetism, Foudations, Theory and Applications, Ed. T.W. Barrett & D.M. Grimes, p. 620-635, 1995.

[46] C.J. Carpenter & R.L. Coren, Teaching electromagnetism in terms of potentials instead of fields, Proceedings of the Second European Conference on "Physics Teaching in Engineering Education", Budapest University of Technology and Economics, http://www.bme.hu/ptee2000/papers/carpent.pdf

[47] G. Cavalleri & E. Tonni, Experimental proof of standard electrodynamics by measuring the self-force on a part of a current loop, Phys. Rev. E 58, p. 2505–2517,1998.

[48] A. K. T. Assis, Comment on "Experimental proof of standard electrodynamics by measuring the self-force on a part of a current loop", Phys. Rev. E 62, 7544, 2000.

[49] P. Graneau & N. Graneau, Electrodynamic force law controversy, Phys. Rev. E 63, 058601, 2001.

[50] G. Cavalleri & E. Tonni, Reply to "Comment on 'Experimental proof of standard electrodynamics by measuring the self-force on a part of a current loop' ", Phys. Rev. E 62, 7545, 2000.

[51] G. Cavalleri, E. Tonni, and & Spavieri, Reply to "Electrodynamic force law controversy", Phys. Rev. E 63, 058602, 2001.

[52] L. Allen, S.M. Barnett & M.J. Padgett (ed), Optical Angular Momentum, Bristol: Institute of Physics Publishing, 2003.

[53] R. Moreau, Magnetohydrodynamics, Kluwer Academic Publishers, 1990.